\documentstyle[aps,epsfig,multicol]{revtex}
\draft
\input epsf
\begin{document}
\title{Nonlinear dynamics of dimers on periodic substrates}
\author{C. Fusco\thanks{Author to whom correspondence should be
addressed. Electronic address: fusco@sci.kun.nl.}, A. Fasolino and T. Janssen}
\address{Department of Theoretical Physics, University of Nijmegen,
Toernooiveld 1, 6525 ED Nijmegen, The Netherlands}

\maketitle

\begin{abstract}

We study the dynamics of a dimer moving on a periodic one-dimensional 
substrate as a function of the initial kinetic energy at zero temperature. 
The aim is to describe, in a simplified picture, the microscopic 
dynamics of diatomic molecules on periodic surfaces,
which is of importance for thin film formation and crystal growth.
We find a complex behaviour, characterized by a variety of
dynamical regimes, namely oscillatory, ``quasi-diffusive'' (chaotic)
and drift motion. Parametrically resonant excitations of internal
vibrations can be induced both by oscillatory and drift motion of the centre 
of mass. For weakly bound dimers a 
chaotic regime is found for a whole range of velocities between two 
non-chaotic phases at low and high kinetic energy. The chaotic features have
been monitored by studying the Lyapunov exponents and the power spectra.
Moreover, for a short-range interaction, the dimer can dissociate due to 
the parametric excitation of the internal motion.

\end{abstract}

\begin{multicols}{2}
\section{Introduction}
\label{sec:intro}

Thin-film growth is a topic of great importance both from the theoretical and
experimental point of view~\cite{Venables}. In order to build a microscopic 
theory of crystal growth, it is fundamental to understand isolated-adatom 
surface diffusion, and indeed a great amount of work has been devoted  
to the monomer case during the last decades~\cite{Risken,Kellog,Feibelman,Ferrando,Guantes}. 
Once individual atoms are adsorbed on a surface, they can meet each
other thus forming dimers. In spite of this very simple morphology, dimers 
display a peculiar and intriguing diffusive behaviour~\cite{Wang,Braun1,Braun2,Montalenti1,Montalenti2,Montalenti3,Bogicevic,Boisvert}.
In this paper we show that the strong nonlinearities arising during the 
motion of dimers on a periodic substrate make dimer dynamics a very complex
phenomenon. Here, we study the problem of a one-dimensional dimer moving on a 
rigid periodic substrate at $T=0$. A one-dimensional model is relevant 
since one-dimensional dimer diffusion occurs in real systems, in particular 
along steps and on channeled $(110)$ metal surfaces~\cite{Kurpick}.
Although some aspects of dimer energetics and dynamics have been recently 
considered (see in particular Refs.~\cite{Braun1,Braun2,Montalenti2}), 
here we focus on the nonlinear microscopic dynamics of this model,
which gives rise to different regimes depending on the initial velocity and 
on the strength of the dimer interaction. 
In particular, chaotic motion with long jumps
is found between the oscillatory behaviour at low initial 
kinetic energy and the drift motion at high initial kinetic energy. 
In all these regimes, situations can be found where the centre of mass (CM)
motion drives excitations of the dimer vibrations which can lead to 
dissociation for short range interatomic interactions. 
Such a complexity in the dynamical behaviour at $T=0$ is relevant 
to understand the thermal diffusion problem. In particular, the chaotic
behaviour which dominates for weakly bound dimers can account for the 
non-trivial dependence of the diffusion constant on the strength of the 
interatomic interaction reported in Ref.~\cite{Braun2}. 

Chaotic motion can occur in nonlinear systems with at least 
three variables. Much studied is the case of systems characterized by a 
single spatial coordinate subjected to an external drive~\cite{Venkatesan,Kim,Kao,Parlitz,Murali}.
Besides, a system of interacting particles can exhibit chaotic motion with
``quasi-diffusive'' features in one dimension~\cite{Strunz,Ou,Kocic}. 
This can happen even without the presence of an external drive, 
as we will show below.
In view of the sinusoidal potential and of the phase space dimension, 
our model bears some resemblance to an undriven double pendulum~\cite{Shinbrot,Levien,Christini}. 
However, in our case, the possibility to perform either oscillatory or 
drift motion leads to the appearance and subsequent disappearance of chaos 
for increasing initial kinetic energy. 

In Sect.~\ref{sec:model} we describe our model. Sect.~\ref{sec:linear} is 
devoted to the discussion of the linearized version of the system, in which
a semi-analytical treatment can be performed. Sect.~\ref{sec:full} analyses
the dynamics of the full system, and in Sect.~\ref{sec:chaos} we
discuss the chaotic properties of the system. 
Some concluding remarks are presented in Sect.~\ref{sec:conclusion}.

\section{Model}
\label{sec:model}

We consider a dimer moving on a periodic one-dimensional substrate at zero 
temperature without damping. 
The particle-substrate interaction is modelled by a periodic 
function $U$:
\begin{equation}
U(x_1,x_2)=U_0(2-\cos(kx_1)-\cos(kx_2))  
\end{equation}
where $x_i$ represents the spatial coordinate of particle $i$ ($i=1,2$), 
$2U_0$ is the potential barrier per particle and $k=2\pi/a$, $a$ being the 
substrate lattice constant.
Most of the results presented here have been obtained using a 
harmonic interatomic potential:
\begin{equation}
\label{e.harmonic}
V(x_1,x_2)=\frac{K}{2}(x_2-x_1-l)^2,
\end{equation}
where $K$ is the force constant and $l$ is the spring equilibrium length.
We also used short range interatomic interactions, such as the Lennard-Jones
(LJ) potential (see Sect.~\ref{sec:LJ}). 

The equations of motion are:
\begin{equation}
\left\{ \begin{array}{ccc}
m\ddot{x}_1 & = & K(x_2-x_1-l)-kU_0\sin(kx_1) \\
m\ddot{x}_2 & = & K(x_1-x_2+l)-kU_0\sin(kx_2)
\end{array} \right.
\end{equation}
where $m$ is the mass of each particle. 
We rescale the variables in the following way:
$$
\tilde{x}_i=kx_i,\quad \tilde{t}=t/\tau,\quad \tilde{U_0}=U_0/E_T,\quad 
\tilde{l}=kl,\quad \tilde{K}=K/(E_Tk^2)  
$$
where $E_T$ is a reference energy and $\tau=[m/(E_Tk^2)]^{1/2}$.
In these units the equations of motion become 
(in the following we will omit the tildes for simplicity)
\begin{equation}
\label{e.motion}
\left\{ \begin{array}{ccc}
\ddot{x}_1 & = & K(x_2-x_1-l)-U_0\sin x_1 \\
\ddot{x}_2 & = & K(x_1-x_2+l)-U_0\sin x_2
\end{array} \right.
\end{equation}
In the CM and relative coordinates frame we have 
\begin{equation}
\label{e.CM}
\left\{ \begin{array}{lll}
\ddot{x}_{CM} & = & -U_0\sin x_{CM}\cos(x_r/2+l/2) \\
\ddot{x}_r & = & -2Kx_r-2U_0\cos x_{CM}\sin(x_r/2+l/2)
\end{array} \right.
\end{equation}
where $x_{CM}=(x_1+x_2)/2$ is the CM coordinate and $x_r=x_2-x_1-l$ is the 
relative coordinate.
We concentrate on the commensurate case in which $l=a=2\pi$, 
i.e. the spring natural length is equal to the period of the substrate 
potential. In this situation the minimum energy configuration does not depend 
on $K$ (namely $x_1=0$ and $x_2=a$ minimize the total potential energy) and 
moreover a linearization around $x_r=0$ offers the possibility to treat the 
problem in a semi-analytical way. 
This has the advantage to give a closer insight on the 
dynamical features of this system. 

We perform molecular dynamics simulations, integrating the equations of 
motion~(\ref{e.motion}) using a velocity-Verlet algorithm, with time step 
$\Delta=10^{-4}$.

\section{Dynamics of the linearized system}
\label{sec:linear}

\begin{figure}
\epsfig{file=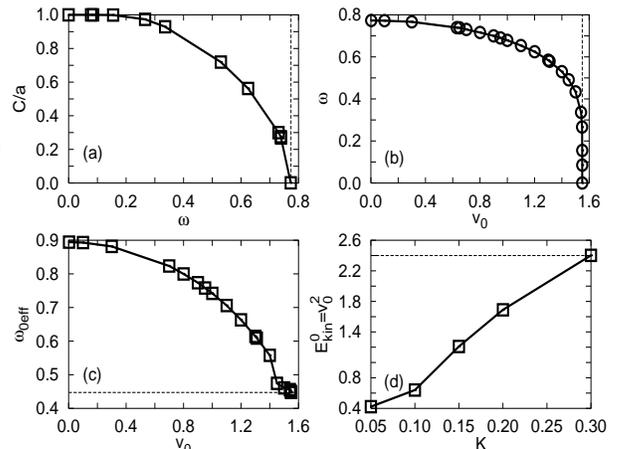,width=8cm,height=6cm}
\caption{Relation between amplitude and frequency of oscillation as given
by the solution of Eq.~\ref{e.motionlin1} (a), between
CM frequency of oscillation and CM initial velocity (b) and between 
the effective stretching frequency and CM initial velocity (c), obtained by 
numerical calculations for $U_0=0.6$ (solid lines). 
The lower borders of instability regions as a function of $K$ are plotted in
(d). The vertical dashed lines in (a) and in (b) 
indicate respectively the frequency of small oscillations 
$\sqrt{U_0}=0.774$ and the velocity corresponding to the threshold 
for drift motion $\sqrt{4U_0}=1.5492$. The horizontal dashed lines in (c) and
(d) correspond respectively to $\omega_{0eff}=\sqrt{2K}$ ($K=0.1$),
which is reached when $v_0\ge \sqrt{4U_0}$, and to the energy threshold for
drift motion $\sqrt{4U_0}$.}
\label{f.freqampinst}
\end{figure}

When $x_r\simeq 0$, as at the beginning of the motion starting from 
equilibrium, we can linearize in $x_r$ the equations of motion~(\ref{e.CM}):
\begin{mathletters}
\begin{eqnarray}
\label{e.motionlin}
\ddot{x}_{CM} & \simeq & U_0\sin x_{CM} \label{e.motionlin1}\\
\ddot{x}_r & \simeq & -2K(1-\frac{U_0}{2K}\cos x_{CM})x_r.\label{e.motionlin2}
\end{eqnarray}
\end{mathletters}
As initial conditions we choose 
$$
\begin{array}{ll}
x_{CM}(0)=x_0=a/2 & \dot{x}_{CM}(0)=v_0 \\
x_r(0)=0      & \dot{x}_r(0)=0.
\end{array}
$$
In this way we give an initial kinetic energy to the dimer at equilibrium
(alternatively one could have chosen to give an initial potential energy, i.e.
$\dot{x}_{CM}(0)=0$ and $x_{CM}(0)\ne a/2$).
The minimum kinetic energy for the CM to get out of the potential is
$v_0^2=4U_0$ if $v_1=v_2=v_0$ ($v_i$ is the initial velocity of particle 
$i$). 
Below this threshold value (namely $v_0<\sqrt{4U_0}$) Eq.~(\ref{e.motionlin1})
coincides with that of a classical pendulum for which the amplitude as a 
function of the period of oscillation is known in terms of an elliptic 
integral~\cite{Berkeley}, as shown in Fig.~\ref{f.freqampinst}(a).
The maximum amplitude of oscillation of the CM is determined by $v_0$. 
In this case 
\begin{equation}
x_{CM}\simeq x_0+\frac{C}{2}\sin(\omega t).
\end{equation}
with $\omega=\omega(v_0)$ (see Fig.~\ref{f.freqampinst}(b)).
This means that
\begin{equation}
\label{e.cos}
\cos x_{CM}\simeq A+B\cos(2\omega t)
\end{equation}
with $A$ and $B$ depending on $v_0$ via $\omega$.
Inserting~(\ref{e.cos}) into the equation of motion
for $x_r$~(\ref{e.motionlin2}) we obtain the equation of a parametric 
oscillator:
\begin{equation}
\label{e.parametric}
\ddot{x}_r=-\omega_{0eff}^2(1+h\cos(2\omega t))x_r
\end{equation}
where $\omega_{0eff}\equiv\sqrt{2K-AU_0}$ and 
$h\equiv BU_0/\omega_{0eff}^2$.
It is worth noting that the stretching frequency of the free dimer 
$\sqrt{2K}$ becomes $\sqrt{2K+U_0}$ in the external potential
(this is true when the CM is fixed at the equilibrium position). 
But in Eq.~(\ref{e.motionlin}) the motion of $x_r$ is further affected by 
the oscillatory behaviour of the CM and its natural frequency changes into
$\omega_{0eff}\equiv\sqrt{2K-AU_0}$.
Conversely, when $v_0>\sqrt{4U_0}$ the CM performs a drift motion, i.e.
\begin{equation}
x_{CM}\simeq x_0+<v_{CM}>t,
\end{equation}
where $<\cdot>$ denotes time averages and $<v_{CM}>$ is the drift CM 
velocity.
By defining $\omega$ as $\omega=<v_{CM}>/2$ the equation for $x_r$ remains in 
the form~(\ref{e.parametric}) with
$\omega_{0eff}\equiv\sqrt{2K}$ and $h\equiv U_0/(2K)$ (i.e. $A=0$ and 
$B=1$).

\begin{figure}
\epsfig{file=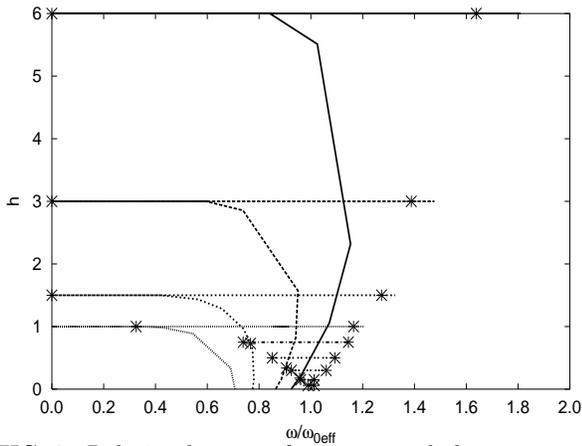,width=8cm,height=6cm}
\caption{Relation between frequency and the parameter h of 
Eq.~(\ref{e.parametric}) for different values of $K$ (from top to bottom:
$K=0.05,0.1,0.2,0.3,0.4,0.6,1,2,5$) and $U_0=0.6$. The region in which the
parametric resonance occurs is bounded by the stars.}
\label{f.instability}
\end{figure}

The CM motion (either oscillatory or drifting), as considered in the 
linearized Eq.~(\ref{e.motionlin}), drives parametrically the internal  motion
of the dimer.
We establish the instability ranges by monitoring
for which values of the initial velocity $v_0$ an exponential
increase of $x_r$ is found. 
The relation between the initial velocity $v_0$ and the frequency
$\omega_{0eff}$ is shown in Fig.~\ref{f.freqampinst}(c) for several values of
$v_0<\sqrt{4U_0}$.
In order to understand the energy threshold for the excitation of parametric
resonances, we plot the lower borders of the instability regions as a function
of $K$ in Fig.~\ref{f.freqampinst}(d) when the total initial kinetic energy 
$E_{kin}^0=v_0^2$ is less than the potential barrier $4U_0$.
In this way we can identify a critical value $K=K_c$ 
above which the parametric resonance can be excited only if the CM 
performs a drift motion (e.g. $E_{kin}^0>4U_0$). 
It turns out that $K_c\simeq 0.3$ for $U_0=0.6$.
Since $\omega_{0eff}$ is determined by $K$, by considering different values
of $K$ we can construct the standard picture for parametric instabilities
relating $h$ to $\omega$. One can recognize the main resonance for 
$\omega=\omega_{0eff}$ (Fig.~\ref{f.instability}). 
The boundaries of the region of instability are 
given by the stars in Fig.~\ref{f.instability}. 
The different curves represent $h$ as a function of $\omega$ 
for different values of $K$. Note that, at fixed $K$, 
$h$ increases when $\omega$ decreases (e.g. when the amplitude of oscillation 
increases and $v_0<\sqrt{4U_0}$), but when the CM overcomes the barrier,
$h$ reaches the constant value $U_0/(2K)$. 
Moreover, the range of frequency in which instability is
observed is larger for smaller values of $K$.
An example of parametric resonance is shown in 
Fig.~\ref{f.dimerdetU_00.6K0.05eta0l1v0.7}(a): 
$x_r$ oscillates and its amplitude increases exponentially.
In Fig.~\ref{f.dimerdetU_00.6K0.05eta0l1v0.7}(b) a blow up of the behaviour 
of $x_r$ and of the drive $x_{CM}$, oscillating at the same frequency, is
also shown.

\section{Dynamics of the full system}
\label{sec:full}

\subsection{Harmonic case}
\label{sec:harmonic}

\begin{figure}
\epsfig{file=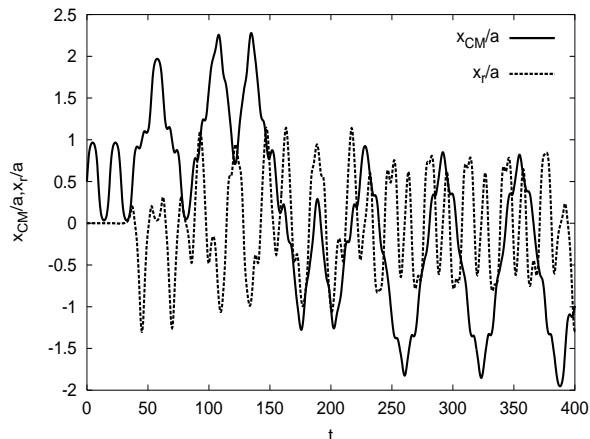,width=8cm,height=6cm}
\caption{Numerical simulations of Eq.~(\ref{e.CM}) for $U_0=0.6$, $K=0.05$ and
$v_0=1.54$. The CM (solid line) and relative motion (dashed line), 
rescaled to the substrate lattice constant $a$, are shown.}
\label{f.dimerdetU_00.6K0.05eta0l1v1.54}
\end{figure}

Now we consider the original system of equations of motion Eq.~(\ref{e.CM}).
The linearization given by Eq.~(\ref{e.motionlin}) allows to understand some 
of the dynamical features of the full system. However, the CM and relative 
motion equations are coupled and this results in a qualitatively different 
behaviour with respect to the simple approximation discussed in 
Sect.~\ref{sec:linear}. 
In particular, we note that the feedback of
$x_r$ on $x_{CM}$ drives the CM out of the instability window found
for the linearized system Eq.~(\ref{e.parametric}).
\begin{figure}
\epsfig{file=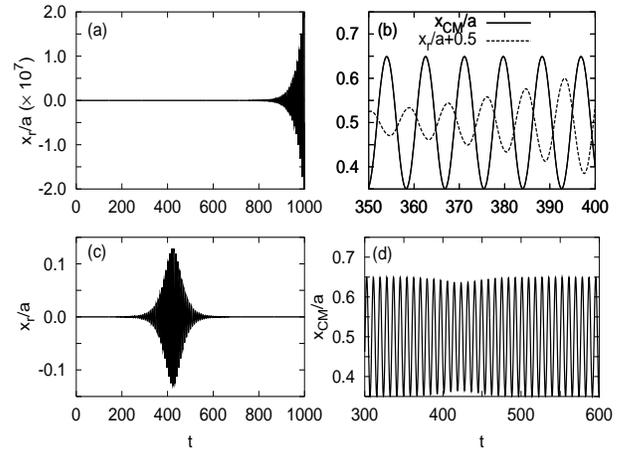,width=8cm,height=6cm}
\caption{Comparison between relative and CM motion of the linearized system 
Eq.~(\ref{e.motionlin}) (upper panel) and those obtained by integrating the 
complete system Eq.~(\ref{e.motion}) (lower panel). The relative motion 
plotted in (a) is the numerical result of the integration of 
Eq.~(\ref{e.parametric}). The CM and relative coordinate of the linearized
equations, which are plotted in (b), oscillate with the same frequency.
The CM of the full system is shown in (d) where we note a decrease of 
amplitude at the point in which the relative motion starts to decrease in 
(c). 
Only the envelope of the rapid oscillations of $x_r$ is visible on the
left panel ((a) and (c)).
The parameters used in the simulation are $U_0=0.6$, $K=0.05$ and 
$v_0=0.7$. All lengths are rescaled to the substrate lattice constant $a$.}
\label{f.dimerdetU_00.6K0.05eta0l1v0.7}
\end{figure}
An example is illustrated in Fig.~\ref{f.dimerdetU_00.6K0.05eta0l1v0.7}, 
where we compare the motion of the complete and of the linearized system. 
As we can see in Fig.~\ref{f.dimerdetU_00.6K0.05eta0l1v0.7}(c), 
the parametric increase of $x_r$ found for the linearized equations is 
followed by a decrease, due to the fact that the feedback of $x_r$ on 
$x_{CM}$ causes a change of the amplitude of the CM
at that point ($t \simeq 400$ in 
Fig.~\ref{f.dimerdetU_00.6K0.05eta0l1v0.7}(d)).
However, when $x_r$ decreases, the instability reappears and 
the relative motion increases again (although it is not shown in the figure). 
The system gets in and out the instability window,
because we are considering a case that is almost at the border.  
Instead, in Fig.~\ref{f.dimerdetU_00.6K0.05eta0l1v1.54} 
we show a case which is in the centre of the region of
instability: we can note that after an initial transient the relative motion 
is always excited, and its behaviour is more irregular so that it is not 
possible to identify a clear unique frequency of oscillation and a unique 
rate of increase.
This behaviour is caused by the shift in position inside the 
instability window which in turn produces a shift in frequency and rate of
increase. Note that the excitation of the internal motion leads to a CM
motion which would not have occurred if the dimer had been rigid. 
In that case the CM would have kept the initial oscillatory behaviour around
the equilibrium position. Here, instead, the internal vibrations play a role 
similar to that of a heat bath and drive the CM away from
the minimum with jumps across one or more potential wells.
In fact, for a non-rigid dimer, it is possible to get
out of the well even if $v_0<\sqrt{4U_0}$. This happens because if the 
internal motion is excited, it is possible that one particle 
remains in the minimum whereas the other reaches the nearest maximum.
In this way the energy balance is:
\begin{equation}
\label{e.energybal}
E_{kin}^0=\frac{1}{2}v_1^2+\frac{1}{2}v_2^2=2U_0+\frac{1}{2}K(a/2)^2
\end{equation}
and if $K$ is sufficiently small the right-hand side is smaller than $4U_0$
(we assume $v_1=v_2=v_0$).
Thus vibrational energy can be effective in overcoming the barrier. 
The resulting motion of the dimer 
(Fig.~\ref{f.dimerdetU_00.6K0.05eta0l1v1.54})
can be characterized as chaotic, as shown later in Sect.~\ref{sec:chaos}.

\begin{figure}
\epsfig{file=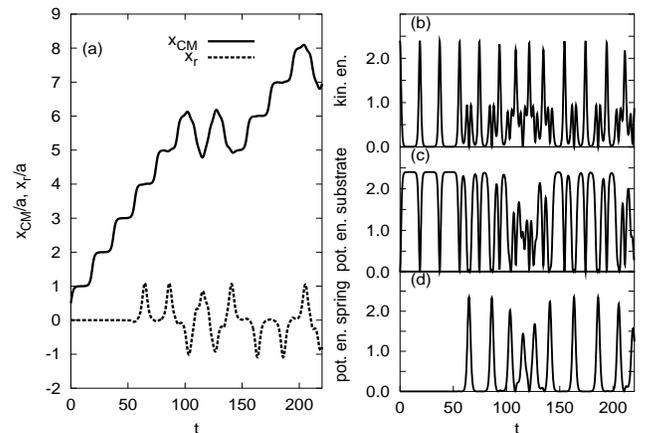,width=9cm,height=6cm}
\caption{Numerical simulations of Eq.~(\ref{e.CM}) for $U_0=0.6$, $K=0.1$ and
$v_0=\sqrt{4U_0}\simeq 1.5492$: (a) CM trajectory (solid line) and 
relative motion (dashed line); (b) kinetic energy;
(c) substrate potential energy; (d) spring potential energy.}
\label{f.dimerdetU_00.6K0.1eta0l1v1.5492}
\end{figure}

It is interesting that the chaotic motion described above occurs at 
velocities below and above the threshold $\sqrt{4U_0}$ for drift motion in 
the linearized system. This is due to a coexistence of 
long jumps with localized motion which persists for a certain range of 
initial energies, as suggested in Ref.~\cite{Guantes}.
In Fig.~\ref{f.dimerdetU_00.6K0.1eta0l1v1.5492}(a) we show the 
case where $v_0=\sqrt{4U_0}$. At the beginning, the CM  performs a step-like 
motion: every time it 
overcomes a barrier it gets stuck for a while in the next minimum before
overcoming the next barrier. In this initial stage $x_r=0$.
After $x_r$ gets excited, this step-like 
motion disappears. The parametric resonance which
one would have expected in this case for the simplified system is not visible
because of the reciprocal influence of CM and relative motion, 
which inhibits the increase of amplitude of $x_r$. 
Note that when the internal 
motion is excited almost all the energy is transferred to the vibrational 
modes, as it can be seen by the corresponding peaks in
Fig.~\ref{f.dimerdetU_00.6K0.1eta0l1v1.5492}(d).

\begin{figure}
\epsfig{file=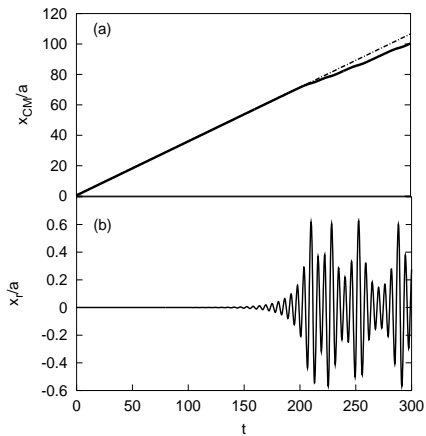,width=9cm,height=6.5cm}
\caption{Numerical simulations of Eq.~(\ref{e.CM}) for $U_0=0.6$, $K=0.5$ and
$v_0=2.5$. In (a) we show the CM motion (solid line) with a linear fit for
$t<200$ (dashed line), while the relative coordinate is plotted in (b).}
\label{f.dimerdetU_00.6K0.5eta0l1v2.5}
\end{figure}

By further increasing the initial kinetic energy the dynamics becomes again
non chaotic, and the CM performs a drift motion with constant velocity unless 
the conditions for parametric excitation given by Eq.~(\ref{e.parametric})
are met. This does not occur for the small values of $K$ considered up to now.
In fact, a dimer 
presents only one characteristic frequency so that conditions for parametric 
excitation are generally met either in the oscillatory or in the drift regime.
The situation would be different for a larger molecule where different 
vibrational modes could be excited for different values of $v_0$.
In Fig.~\ref{f.dimerdetU_00.6K0.5eta0l1v2.5} we show one situation for large
$K$ where the drift CM motion excites the internal motion through
a parametric resonance for drift velocity twice 
the dimer natural stretching frequency $\omega_{0}=\sqrt{2K}$.  
When the relative motion acquires a large amplitude, deviations from the 
linear behaviour of $x_{CM}$ are observed.

\begin{figure}
\epsfig{file=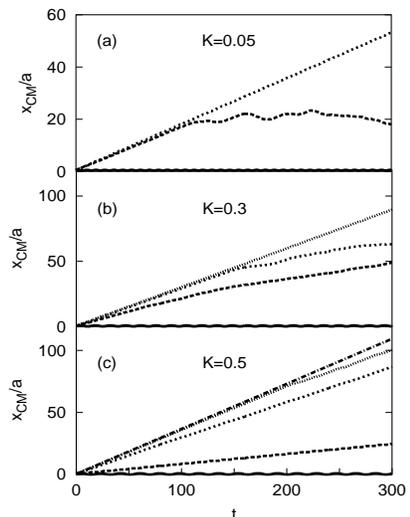,width=10cm,height=7cm}
\caption{Numerical simulations of Eq.~(\ref{e.CM}) for $U_0=0.6$, three 
values of $K$ and several values of $v_0$. The CM motion is plotted for 
$K=0.05$ (a), $K=0.3$ (b) and $K=0.5$ (c). The different curves in each graph
are obtained with different initial velocities. From bottom to top in 
each graph $v_0=0.63,1.65,1.68$ (a), $v_0=1.5,1.85,2.15,2.2$ 
(b), $v_0=1.54,1.55,2.2,2.5,2.55$ (c). Note the deviation from linear 
behaviour in (c) for $v_0=2.5$ (see Fig.~\ref{f.dimerdetU_00.6K0.5eta0l1v2.5}
and text).}
\label{f.xcmdetU_00.6Kalleta0l1vall}
\end{figure}

In Fig.~\ref{f.xcmdetU_00.6Kalleta0l1vall} we summarize for three values 
of $K$ the effect of different $v_0$ (at fixed $U_0$) on the 
CM  motion. 
Increasing $v_0$, a complex transition from oscillatory regular motion 
to chaotic motion and then to a drift regime can take place depending on the
value of $K$.
For $K=0.05$ (Fig.~\ref{f.xcmdetU_00.6Kalleta0l1vall}(a)), 
resonant excitation of the internal motion occurs 
for $v_0>0.65$ as given in Fig.~\ref{f.freqampinst}(d). 
Above this value, first a regime with 
recursive excitation of $x_r$, as in 
Fig.~\ref{f.dimerdetU_00.6K0.05eta0l1v0.7}(c), takes place so that 
$<x_{CM}>=a/2$ and  $<v_{CM}>=0$. 
At $v_0>1.3$ the escape from the well 
described by Eq.(~\ref{e.energybal}) becomes possible. The resulting (chaotic)
behaviour of $x_{CM}$ in this regime is shown by the dashed line in 
Fig.~\ref{f.xcmdetU_00.6Kalleta0l1vall}(a) for $v_0=1.65$. 
The CM motion in this regime
behaves as $<x_{CM}^2>\simeq t^{\alpha}$ ($1<\alpha<2$) with
$<v_{CM}>\simeq 0$, i.e. it is 
``quasi-diffusive'', that is to say neither purely diffusive nor ballistic.
This behavior extends up to $v_0 < 1.68$ , i.e. well above the "threshold" 
$\sqrt{4U_0}$. Above, a drift motion  $x_{CM}(t)\simeq x_0+<v_{CM}>t$ occurs. 
For a larger $K=0.3$ (Fig.~\ref{f.xcmdetU_00.6Kalleta0l1vall}(b))
the quasi-diffusive motion starts occurring  at 
values of $v_0\simeq \sqrt{4U_0}$ up to $v_0=2.2$, where drift motion 
is recovered. 
Lastly for $K=0.5$ (Fig.~\ref{f.xcmdetU_00.6Kalleta0l1vall}(c)), there is
no chaotic regime. The drift motion starts at $v_0=\sqrt{4U_0}$ and 
deviations only occur for a narrow range of velocities where $x_r$ becomes 
parametrically excited, as shown in Fig.~\ref{f.dimerdetU_00.6K0.5eta0l1v2.5}.
 
We may estimate from 
Eq.~(\ref{e.energybal}) the critical $K$ value above which the internal 
motion is not effective in making the CM overcome the potential barrier for 
$v_0<\sqrt{4U_0}$. 
Namely
\begin{equation}
2U_0+\frac{1}{2}K_c\pi^2=4U_0,
\end{equation}
and for $U_0=0.6$ we find $K_c\simeq 0.35$.
As a consequence for $K>K_c$ no chaotic motion is found.

\subsection{Lennard-Jones case}
\label{sec:LJ}

\begin{figure}
\epsfig{file=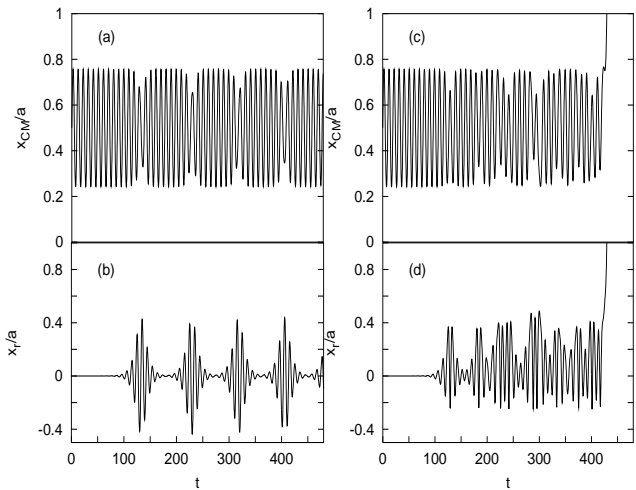,width=8.5cm,height=6.7cm}
\caption{Numerical simulations of the equation of motion of a dimer for 
harmonic (a),(b) and LJ interactions (c),(d). 
The CM and relative motion are shown. The parameters are $U_0=0.6$, $K=0.05$, 
$v_0=1.12$ for the harmonic potential. The parameters of the 
LJ have been chosen according to Eq.~(\ref{e.sigmaepsilon}):
$\sigma\simeq 5.598$ and $\epsilon\simeq 0.0274$.}
\label{f.ljdimerdetU_00.6K0.05eta0l1v1.12}
\end{figure}

Now we consider the effect of replacing the harmonic 
interaction~(\ref{e.harmonic}) with a finite-range potential. 
As a simple choice, we take the LJ potential, given by
\begin{equation}
V_{LJ}(r)=4\epsilon\left[\left(\frac{\sigma}{r}\right)^{12}-
\left(\frac{\sigma}{r}\right)^6\right]
\end {equation}
with $r\equiv |x_2-x_1|$ and a cutoff at $r=2.5\sigma$.
To recover the harmonic interaction close to the minimum,  
we impose the equilibrium distance to be 
equal to the spring equilibrium length and the second derivative of $V_{LJ}$
to be equal to the spring constant, namely:
\begin{equation}
\left\{\begin{array}{ccccc}
r_{min} & = &\sqrt[6]{2} \sigma & = &l \\
\frac{d^2V_{LJ}}{dr^2}\Big|_{r=r_{min}} & = &\frac{24\epsilon}{\sigma^2}
\left[\frac{26}{\sqrt[3]128}-\frac{7}{\sqrt[3]16}\right] &= &K
\end{array} \right.
\end{equation}
whence
\begin{equation}
\label{e.sigmaepsilon}
\left\{\begin{array}{ccc}
\sigma & = & \frac{l}{\sqrt[6]2}\\
\epsilon & = & \frac{Kl^2}{72}
\end{array} \right. 
\end{equation}
At variance with harmonic interactions, a finite-range potential allows 
dissociation of particles. This is clearly seen in 
Fig.~\ref{f.ljdimerdetU_00.6K0.05eta0l1v1.12}, where the time behaviour of 
the CM and relative motion is plotted comparing the harmonic and 
LJ potentials. 
The CM is the same in the two cases when $x_r=0$, 
i.e., at the beginning of the motion. Then, as the relative motion starts to 
increase, $x_r$ given by LJ is found very similar to the
harmonic $x_r$, but then the amplitude of the oscillations due to the 
LJ potential becomes larger. At $t\simeq 420$  the two 
particles dissociate and $x_r$ starts to increase very fast since only one 
particle moves.

\begin{figure}
\epsfig{file=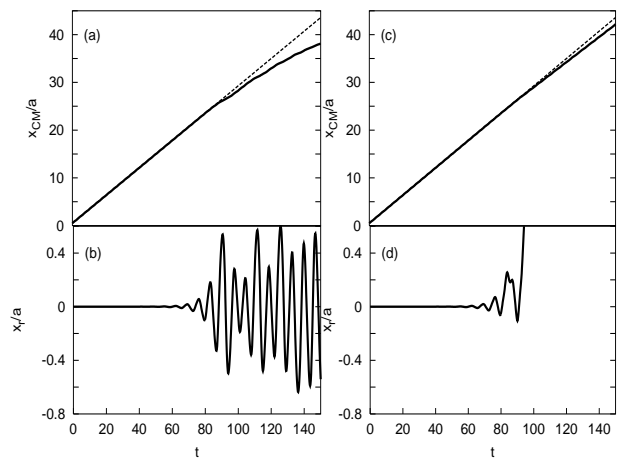,width=8.3cm,height=6.3cm}
\caption{Numerical simulations of the equation of motion of a dimer for 
harmonic (a),(b) and LJ interactions (c),(d). The CM and relative 
motion are shown. The dashed lines in (a) and (c) are linear fits to $x_{CM}$
for $t<80$. The parameters are $U_0=0.6$, $K=0.4$ and $v_0=2.15$ for the 
harmonic potential. The parameters of the LJ have been chosen
according to Eq.~(\ref{e.sigmaepsilon}): $\sigma\simeq 5.598$ and 
$\epsilon\simeq 0.219$.}
\label{f.ljdimerdetU_00.6K0.4eta0l1v2.15}
\end{figure}

In Fig.~\ref{f.ljdimerdetU_00.6K0.4eta0l1v2.15} we show a similar process for 
the  case where the CM performs a drift motion. 
We observe that the increase of amplitude of $x_r$ occurs approximately 
at the same time for  both harmonic and LJ interactions. As just noted above, 
at this point a departure from the linear behaviour of the CM takes place. 
While large oscillations persist in the harmonic $x_r$, 
breaking of the interparticle bond is found in the LJ case. This shows that 
the resonant excitation of internal vibrations could be effective in leading 
to dissociation of molecular bonds.

\section{Chaotic dynamics}
\label{sec:chaos}

\begin{figure}
\epsfig{file=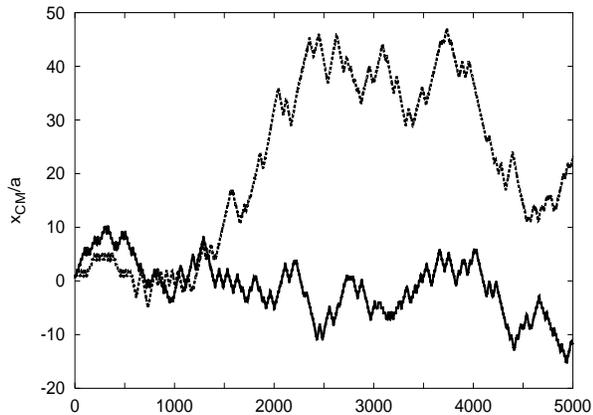,width=8cm,height=6cm}
\caption{CM motion for $U_0=0.6$, $K=0.1$ and $v_0=\sqrt{4U_0}\simeq 1.5492$,
starting from different initial conditions: $x_1(0)=0$, $x_2(0)=a$ 
(solid line) and $x_1(0)=10^{-6}$, $x_2(0)=a+10^{-6}$ (dashed line).}
\label{f.xcmdetU_00.6K0.1eta0l1v1.5492twopi-pmtwopi}
\end{figure}

The dynamics described in Sect.~\ref{sec:harmonic} shows very complex 
features, in spite of the simplicity of our model. The quasi-diffusive, 
irregular motion found for small values of $K$, as in 
Fig.~\ref{f.dimerdetU_00.6K0.05eta0l1v1.54}, resembles characteristics 
peculiar to a chaotic regime. This is confirmed by looking at the 
temporal evolution of two trajectories starting at infinitesimally distant
points. For example, Fig.~\ref{f.xcmdetU_00.6K0.1eta0l1v1.5492twopi-pmtwopi} 
shows two long CM trajectories with initial spatial conditions differing from 
$10^{-6}$. The behaviour of the CM is unpredictable and the trajectories 
diverge for the entire simulation time.
This is a qualitative signature of chaotic dynamics.
In order to characterize more quantitatively the chaotic motion, we 
have numerically computed the Lyapunov exponent, which measures
the rate of divergence of nearby trajectories (see~\cite{Eckmann1,Eckmann2}):
\begin{equation}
\delta x(t)\sim \delta x(0){\rm e}^{\lambda t},
\end{equation}
where $\delta x(t)$ denotes the separation between nearby trajectories and
$\lambda$ is the Lyapunov exponent. 

\begin{figure}
\epsfig{file=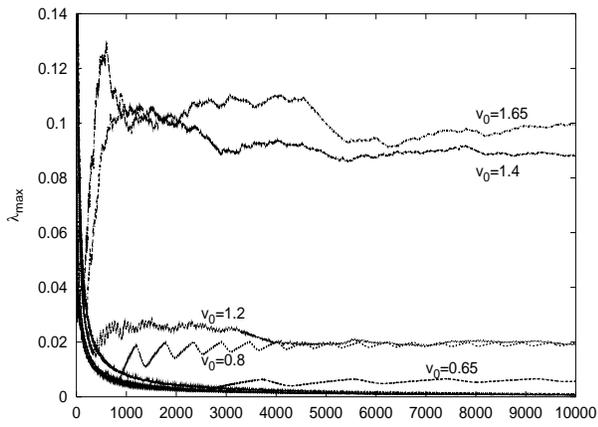,width=8cm,height=6cm}
\caption{Temporal behaviour of the maximal Lyapunov exponent for $U_0=0.6$,
$K=0.05$ and different initial velocities $v_0$, which are reported beside
each curve, except for the two curves that saturate at zero, which correspond
to $v_0=0.63$ and $v_0=1.68$.}
\label{f.lyapunov}
\end{figure}

In an $n-$ dimensional phase space $n$ Lyapunov exponents can be calculated, 
but we limit ourselves to the computation of the maximal Lyapunov exponent 
$\lambda_{max}$, which is sufficient to signal the occurrence of
chaos. If $\lambda_{max}>0$ the motion is unstable and chaos may occur, 
while if $\lambda_{max}<0$ we have a regular stable motion ($\lambda_{max}=0$ 
corresponds to a stable quasi-periodic motion).
We show $\lambda_{max}$ as a function of time for a small value of $K$ 
($K=0.05$ and $U_0=0.6$) in Fig.~\ref{f.lyapunov} for different $v_0$. 
The saturation values of the different curves give a measure of the 
corresponding maximal Lyapunov exponent. 
We note that for low values of $v_0$ ($v_0=0.63$ in the figure)
$\lambda_{max}=0$, meaning that the motion in this range is regular: $x_{CM}$
oscillates periodically and $x_r\simeq 0$. When the internal motion starts to
be excited ($0.65 \le v_0<1.4$) $\lambda_{max}$ jumps to a positive small 
value ($\lambda_{max}\simeq 0.02$), signalling that a weak chaotic dynamics 
is induced by the relative coordinate.
For larger values of $v_0$ ($1.4\le v_0< 1.68$), as the CM gets out of the 
minimum of the potential well and performs an irregular motion of the type shown in Fig.~\ref{f.dimerdetU_00.6K0.05eta0l1v1.54},  
the magnitude of $\lambda_{max}$ suddenly increases of about one order of 
magnitude ($\lambda_{max}\simeq 0.1$), but jumps again discontinuously to
zero when $v_0$ is high enough for the CM to perform a drift motion and 
$x_r\simeq 0$. In this way, we have a complex transition from non-chaotic
to chaotic motion and viceversa as a function of the initial velocity. 
This behaviour is different from that of the double pendulum, where the 
non-chaotic regime is not recovered for large initial velocities~\cite{Shinbrot,Levien}.

\begin{figure}
\epsfig{file=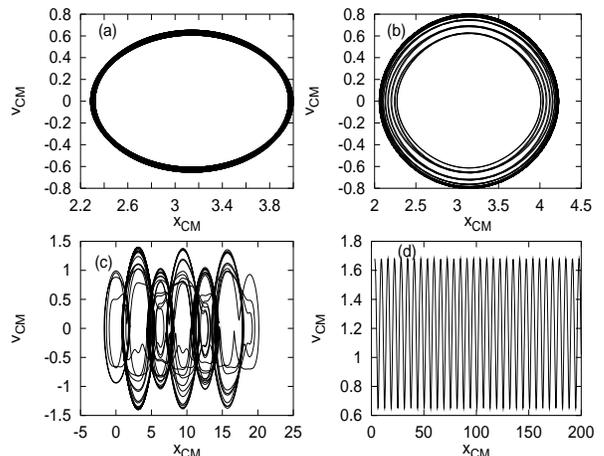,width=8.3cm,height=6.3cm}
\caption{Phase space plot projected on the ($x_{CM},v_{CM}$) plane for 
$U_0=0.6$, $K=0.05$ and four values of $v_0$: $v_0=0.63$ (a), $v_0=0.8$ (b),
$v_0=1.4$ (c) and $v_0=1.68$ (d).}
\label{f.phasespace}
\end{figure}

As a further indicator of such a dynamical behaviour we have plotted the 
phase space projected on the $(x_{CM},v_{CM})$ plane in 
Fig.~\ref{f.phasespace}, for four different initial velocities.
The phase plot in (a) is a simple closed loop corresponding to a regular 
oscillatory
motion where $\lambda_{max}=0$. As the initial velocity increases more 
complex features appear: in the weak chaotic regime (b) extra loops are
present, while the phase plot in (c) becomes very much folded and irregular.
The regular dynamics is restored again in (d), where there is a drift motion
of the CM, with $v_{CM}$ oscillating around the drift velocity.

\begin{figure}
\epsfig{file=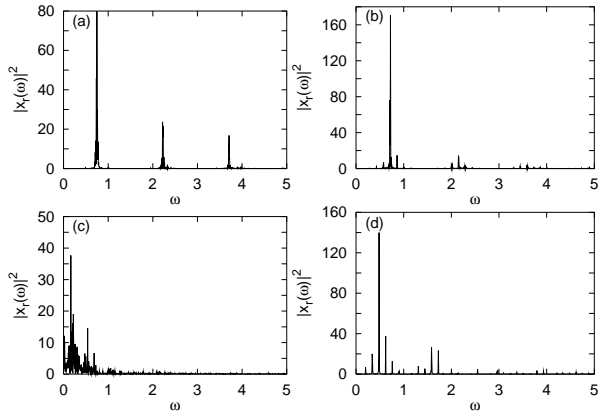,width=8cm,height=5.7cm}
\caption{Power spectra of the relative coordinate $|x_{r}(\omega)|^2$ for 
$U_0=0.6$, $K=0.05$ and  four values of $v_0$: $v_0=0.63$ (a), $v_0=0.8$ (b),
$v_0=1.4$ (c) and $v_0=1.68$ (d).}
\label{f.spectra}
\end{figure}

Power spectrum analysis is usually considered as an additional effective 
method to detect chaos. We have calculated the power spectra of $x_r$ by 
using a fast Fourier transform and we show them in Fig.~\ref{f.spectra} 
for the same values of $U_0$, $K$ and $v_0$ as in Fig.~\ref{f.phasespace}.
The power spectrum for the regular motion (a) is smooth and has few peaks,
at $\omega_{osc},3\omega_{osc},5\omega_{osc},...$, i.e. the harmonics 
expected for a parametric oscillator. In (b) each peak broadens,
developing further lateral features. For the most chaotic motion (c) the 
power spectrum becomes very irregular with a large number of peaks. This
chaoticity disappears for higher velocities (d), where the motion is regular
and the power spectrum is again smooth with a large peak at $\omega=\sqrt{2K}$
corresponding to the dimer stretching frequency.

\section{Conclusions and discussion}
\label{sec:conclusion}

In this paper, we have studied a one-dimensional model of a dimer moving 
on a rigid periodic substrate.
We have shown how the nonlinear effects due to the substrate potential 
and to the interplay between the CM and relative motion
can be relevant in determining the peculiar characteristics of the dimer 
motion. 
A complex dynamical behaviour is found as a function of the initial kinetic 
energy with the occurrence of resonant instabilities and chaotic motion.
In particular, for weakly bound dimers a 
chaotic regime is found for a whole range of velocities between two 
non-chaotic phases at low and high kinetic energy.
We have characterized this 
chaotic regime by studying the Lyapunov exponents and power spectra. 
Moreover, we have shown that if more realistic, finite-range interactions 
are considered, dimer dissociation can be induced via this mechanism by 
choosing appropriate initial velocities. 

Although our model neglects thermal fluctuations, we can try to make some 
qualitative predictions concerning the effect of a finite substrate 
temperature, modelled by stochastic forces and a damping term, on the
diffusive dynamics. The introduction of these effects could smear out all
the deterministic effects. The stable periodic orbits become attracting 
centres via dissipation, so that 
the regular and chaotic motions would be only transient.
On the other hand, temperature should be effective to provide energy to 
escape from an attractor, giving rise to a diffusive motion. 
Nevertheless, we expect that for small temperatures and small friction 
coefficients, the thermal equilibration time should be bigger than the 
equilibration time due to the deterministic chaotic dynamics~\cite{Guantes}. 
Thus, under such circumstances, the effects explained in this paper could be 
significant also for real systems at finite temperature at short time scales.
In particular, diffusion should be highly promoted for weakly bound dimers 
for which we found the chaotic features. 
Indeed, as reported in Ref.~\cite{Braun2}, the 
diffusion coefficient decreases by about an order of magnitude with respect
to the non-interacting case $K=0$, when the elastic constant is increased 
from $K=0$ to $K=0.25$, at least for small values of the damping and the 
temperature. Moreover, signatures of the chaotic regime can be recognized
in the jump length distribution: jumps of the adsorbate over many lattice 
parameters are predicted by theoretical models 
(see for example~\cite{Montalenti2,Braun3}) and observed experimentally 
(see~\cite{Schunack}). 
A preliminary study of the same model at $T\ne 0$ shows that the 
coexistence  of localized and unbounded motion of the type shown in 
Fig.~\ref{f.phasespace}(c),is present up to temperatures a few times higher
than the potential barrier $2U_0$.
The relation between chaotic deterministic diffusion and 
stochastic thermal diffusion is an important topic currently under 
study~\cite{Guantes,Fusco}.
This represents a connection between the behaviour 
of our simplified model and the one of more realistic systems. 
Therefore we believe that, beside their intrinsic 
interest, our results can be of importance to understand the dynamical 
behaviour of dimers moving onto real surfaces.

{\em Note added in proof}. After completion of this manuscript we have noted 
the recent paper A.~S. Kovalev, A.~I. Landau, Low Temp. Phys. \textbf{28},
423 (2002) presenting related numerical results of diffusive dimer dynamics.

\begin{acknowledgements}

This work was supported by the Stichting Fundamenteel Onderzoek der Materie
(FOM) with financial support from the Nederlandse Organisatie voor 
Wetenschappelijk Onderzoek (NWO).

\end{acknowledgements}

\end{multicols}

\end{document}